\documentclass[pra, twocolumn, floatfix]{revtex4}
\usepackage{graphicx}
\usepackage{color}
\usepackage{amsmath, amsfonts, amssymb, bm}
\usepackage{braket}
\usepackage{tikz}
\usepackage{comment}
\usetikzlibrary{arrows}
\usetikzlibrary{decorations.pathmorphing}
\tikzset{snake it/.style={decorate, decoration=snake}}

\begin{document}
\title{Influence of nuclear motion on resonant two-center photoionization}
\author{F. Gr\"ull}
\author{A. B. Voitkiv}
\author{C. M\"uller}
\affiliation{Institut f\"ur Theoretische Physik I, Heinrich Heine Universit\"at D\"usseldorf, Universit\"atsstr. 1, 40225 D\"usseldorf, Germany}
\date{\today}
\begin{abstract}
Photoionization of an atom in the presence of a neighboring atom of different species  is studied. The latter first
undergoes resonant photoexcitation, leading to an autoionizing state of the diatomic system. Afterwards, the
excitation energy is transferred radiationlessly via a two-center Auger process to the other atom, causing
its ionization. Assuming a fixed internuclear distance, it has been predicted theoretically that this indirect
ionization pathway can strongly dominate over the direct photoionization. Here, we extend the theory of
resonant two-center photoionization by including the nuclear motion in van-der-Waals molecules. An analytical
formula is derived reflecting the influence of molecular vibrational dynamics on the relative strength of the
two-center channel. For the specific example of Li-He dimers we show that
the two-center autoionizing resonances are split by  the nuclear motion into multiplets, with the resonance lines reaching a comparable level of enhancement over direct
photoionization as is obtained in a model based on spatially fixed nuclei. 
\end{abstract}
\maketitle

\section{Introduction}
Studies on photoionization processes and other photo-induced breakup reactions in atoms and molecules have always helped deepening our understanding of the structure and dynamics of matter on a microscopic scale. Due to well-defined transfers of energy and momentum during photoabsorption, such processes can provide detailed information on the underlying interaction mechanism as well as the involved atomic system. Accurate and comprehensive tests of corresponding theoretical models are nowadays enabled  by kinematically complete experiments \cite{Ullrich}. 

Within the multitude of different photoionization processes, electron-electron correlations often play an integral part. A well known example is resonant photoionzation, where photoexcitation creates an autoionizing state which subsequently leads to ionization via Auger decay. 
This mechanism can also be generalized to systems consisting of two (or more) atoms. In this case, the resonant excitation of one atom leads -- via
interatomic electron-electron correlations -- to a radiationless energy transfer to a neighbour atom, resulting in its ionization. This decay mechanism is a two-center version of the (usually intraatomic) Auger effect and also well-known as interatomic Coulombic decay (ICD) \cite{ICD,ICDres,ICDrev}. It has been experimentally observed in a variety of systems, e.g. in noble gas dimers \cite{dimers}, clusters \cite{ICDcluster} and water molecules \cite{ICDwater} after photoabsorption. ICD can also be triggered by electron impact which has been demonstrated in various noble-gas systems for a range of impact energies \cite{Grieves,Dorn, Lanzhou, Lanzhou2}.

Stabilization through ICD is included in resonant two-center photoionization (2CPI) which represents
the generalization of resonant photoionization to diatomic systems \cite{2CPI1,2CPI2}. The process describes the ionization of an
atom $A$ via photoexcitation of a neighbouring atom $B$ of different species. This way, an autoionizing two-center state is formed which can decay via a two-center Auger decay, that is through ICD. 2CPI has theoretically been studied in systems of two atoms whose internuclear separation was assumed to be fixed. Under suitable conditions it was shown to dominate over direct (i.e. single-center) photoionization
of atom $A$ by several orders of magnitude. As specific example, a diatomic system of Li and He was considered \cite{2CPI1,2CPI2}. Related theoretical studies (see also \cite{Perina}) have treated two-center single-photon
double ionization \cite{Alex}, strong-field multiphoton ionization \cite{Jacqui} and electron-impact ionization \cite{gruell}
in heteroatomic systems as well as two-photon single-ionization in homoatomic systems \cite{Mueller2011}.

Experiments on two-center processes are mostly performed on van-der-Waals molecules or clusters which
are characterized by small binding energies and large bond lengths. In particular, experimental studies on
2CPI were carried out on He-Ne dimers \cite{2CPIexp,Mhamdi18} and on Ne-Ar clusters \cite{Hergenhahn} using synchrotron radiation to induce
the 1s-3p transition in He at about 23 eV and the 2p-3s transition in Ne at about 17 eV, respectively. Both experiments found strong enhancements of the photoelectron yield as compared with the direct ionization channels of Ne or Ar, respectively. 

In these experiments, the internuclear distance between
the two atoms is not fixed but varies due to the vibrational
molecular motion. Therefore, the original theory of 2CPI \cite{2CPI1}
needs to be further developed to account for the nuclear
motion as well. Of particular interest is the influence of the
latter on the total cross section of 2CPI and its relative
strength in comparison with the direct photoionization.
It is worth mentioning that angular distributions of electrons
emitted via 2CPI from He-Ne dimers have been calculated in
\cite{Mhamdi18}, including their vibrational level structure (for
a similar consideration of ICD in He$_2$, see \cite{Mhamdi20}).
Besides, we note that 2CPI has recently been studied in slow
atomic collisions where the internuclear distance is not fixed
either but subject to the relative atomic motion which was
assumed to follow a straight-line trajectory \cite{Voitkiv}.

\begin{figure}[t]
\begin{center}
\includegraphics[width=0.47\textwidth]{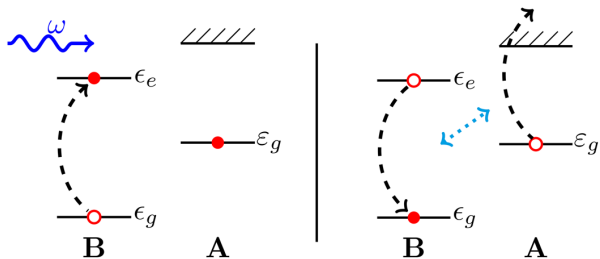}
\end{center}
\vspace{-0.5cm} 
\caption{Scheme of resonant two-center photoionization 2CPI. A two-center autoionizing state is created via photoexcitation of atom $B$ (left). Subsequently, the latter transfers the excitation energy radiationlessly to atom $A$, leading to its ionization via two-center Auger decay (right).}
\label{figure1}
\end{figure} 
In the present paper, we expand the theoretical description of 2CPI by incorporating the nuclear motion in
weakly bound heteroatomic dimers. This way,corresponding treatments for ICD \cite{Moiseyev, Jabbari} are extended
by fully including the photoexcitation step. While 2CPI in van-der-Waals molecules is a rather complex process,
we try to keep its treatment as analytical as possible in order to facilitate the physical
interpretation of the results. Our approach allows us to examine the changes, which originate from the
nuclear motion, to the predictions on 2CPI for fixed internuclear separations \cite{2CPI1,2CPI2}. Besides, a comparison
between 2CPI and the direct photoionization in a dimer is drawn and a formula for the ratio of the
corresponding cross sections is obtained.

Our paper is organized as follows: In Sec. II we present our theoretical considerations of 2CPI.
Applying the Born-Oppenheimer approximation, we first treat the (fast) electronic transitions
involved in 2CPI at fixed internuclear distance \cite{Fussnote2} and, afterwards, include the (slow) nuclear motion
in the potential formed by the Coulomb and van-der-Waals interactions between the atoms $A$ and $B$.
An expression for the 2CPI cross section in a weakly bound dimer is derived which contains transition
matrix elements between the relevant molecular vibrational states. In Sec. III we apply our theory to 2CPI
in Li-He dimers. The influence of the vibrational nuclear motion is demonstrated and its physical implications
are discussed. Concluding remarks are given in Sec. IV. 

Atomic units (a.u.) will be used throughout unless
otherwise stated. The Bohr radius is denoted by $a_{0}$.

\section{Theory of resonant two-center photoionization}
\label{sec:theory}
As in \cite{2CPI1}, we describe the process of 2CPI in a simple system of two atomic centers $A$ and $B$. The geometry of the system, in which both atoms are initially in their ground states, is characterized by the linking vector ${\bf R}$ between the nuclei. Within the Born-Oppenheimer picture, electronic and nuclear processes can be treated separately. The fast electronic transitions can be calculated for fixed interatomic distances $R$, whereas the slow nuclear motion can be considered afterwards.
We note that $R$ has to be sufficiently large (covering at least several Bohr radii) in order to apply the approximations used in the following. 
\subsection{Electronic transitions}
\label{sec:general}
To begin with, we calculate the process treating the two atoms individually and incorporate molecular interactions in the following sections.
The nuclei which are assumed to be at rest carry the corresponding charges $Z_{A}$ and $Z_{B}$.
We take the position of $Z_{A}$ as the origin and denote the coordinates of the  nucleus $Z_{B}$, the electron associated with $A$ and the electron associated with $B$ by $\bf R$, $\bf r$ and ${\bf r'}={\bf R}+{\boldsymbol \xi}$ respectively, where $\boldsymbol \xi$ is the position of the electron of atom $B$ relative to the nucleus $Z_{B}$.

The considered process involves two steps, the photoexcitation of atom $B$ and the subsequent radiationless energy transfer leading to the ionization of $A$. Therefore, we need to define initial, intermediate and final configurations of the two electrons associated with $A$ and $B$, which are illustrated in Fig. \ref{figure1}. Within this basic consideration, we restrict the expressions to one active electron on each center included in the transition. Therefore, we obtain the following states:
(I) The initial state $\Phi_{g,g}=\varphi_{g}({\bf r})\chi_{g}({\boldsymbol \xi})$, with total energy $E_{g,g}=\varepsilon_{g}+\epsilon_{g}$  has both electrons of $A$ and $B$ in their ground state. (II) In the intermediate state $\Phi_{g,e}=\varphi_{g}({\bf r})\chi_{e}(\boldsymbol \xi)$
with total energy $E_{g,e}=\varepsilon_{g}+\epsilon_{e}$, the electron of $B$ has been excited by the electromagnetic field and the electron in $A$ remains in its ground state. (III) The final state  $\Phi_{{\bf k},g}=\varphi_{{\bf k}}({\bf r})\chi_{g}(\boldsymbol \xi)$
with total energy $E_{{\bf k},g}=\varepsilon_{k}+\epsilon_{g}$ and $\varepsilon_{k}=\frac{k^{2}}{2}$  consists of the electron from $A$, which  has been emitted into the continuum with asymptotic momentum $\bf k$ and the electron of $B$, which has returned to its ground state.

In order to ionize atom $A$ in a two-center process including atom $B$, the ionization potential $I_{A}=|\varepsilon_{g}|$ has to be smaller than the energy difference $\omega_{B}=\epsilon_{e}-\epsilon_{g}$ of the electronic transition in atom $B$. The probability amplitude is calculated via the time-dependent perturbation theory,
\begin{eqnarray}
\label{eq: S2}
S^{(2)}&=&-\int\limits_{-\infty}^{\infty}dt\, \mathcal{V}^{AB}({\bf k},{\bf R})e^{-i(E_{g,e}-E_{k,g})t}\nonumber\\&&\times\int\limits_{-\infty}^{t}dt'\, \mathcal{W}^{B}(\omega)e^{-i(E_{g,g}+\omega-E_{g,e})t'}.
\end{eqnarray}
The matrix elements are given by
\begin{eqnarray}
\mathcal{V}^{AB}({\bf k},{\bf R})&=&\langle \Phi_{{\bf k},g}|\hat{V}_{AB}({\bf R})|\Phi_{g,e}\rangle\\
\mathcal{W}^{B}(\omega)&=&\langle\Phi_{g,e}|\hat{W}_{B}(\omega)|\Phi_{g,g}\rangle .
\end{eqnarray}
The operator $ \hat{W}_{B}e^{-i\omega t }$ induces the photoexcitation of atom $B$.
The external field interacting with the atomic centers is described as a classical field with linear polarization. Applying dipole approximation simplifies the employed vector potential:
\begin{equation}
 {\bf A}(t)= {\bf A}_{0}\cos(\omega t).
 \end{equation}
 The direction of polarization is set along the $z$-axis, ${\bf A}_{0}= A_{0}{\bf e}_{z}$. Hence, the interaction $\hat{W}_{\text{B}}$ reads:
  \begin{equation}
  \hat{W}_{B}=\frac{{\bf A}_{0}\cdot{\bf \hat{p}}_{{\bf \xi}}}{2c}.
  \end{equation}
 The interatomic interaction $\hat{V}_{AB}$ causes the energy transfer between $A$ and $B$ resulting in the ionization of $A$. The excited state in atom $B$ decays in a dipole-allowed transition. We neglect retardation effects and obtain
\begin{equation}
\label{eq: VAB}
\hat{V}_{AB}({\bf R})=\frac{{\bf r}\cdot{\boldsymbol \xi}}{R^{3}}-\frac{3({\bf r}\cdot{\bf R})({\boldsymbol \xi}\cdot{\bf R})}{R^{5}}.
\end{equation}
Integration of Eq. \eqref{eq: S2} leads to
\begin{eqnarray}
\label{eq: S2-delta}
S^{(2)}=-2\pi i \delta(E_{{\bf k},g}-E_{g,g}-\omega)\frac{\mathcal{V}^{AB}({\bf k},{\bf R})\mathcal{W}^{B}(\omega)}{\Delta+\frac{i}{2}\Gamma},
\end{eqnarray}
where $\Delta=\epsilon_{g}+\omega-\epsilon_{e}$ is the energy detuning. The resonance width $\Gamma$ accounts for the instability of the excited state of atom $B$ and includes the radiative width $\Gamma_{\text{rad}}$ and the two-center Auger width $\Gamma_{\text{aug}}$  \cite{Fussnote}
\begin{equation}
\Gamma=\Gamma_{\text{rad}}+\Gamma_{\text{aug}}.
\end{equation}
The radiative decay rate $\Gamma_{\text{rad}}$ describes the decay of the excited state of atom $B$ via spontaneous emission of a photon. The two-center Auger width $\Gamma_{\text{aug}}$ describes radiationless transfer of the transition energy in atom $B$ to atom $A$, leading to its ionization. Accordingly, the widths are obtained from the equations
 \begin{align}
  \label{eq: Gamma}
  \Gamma_{\text{rad}}&=\frac{4 \omega_{B}^{3}}{3 c^{3}}\left|\left\langle\chi_{g}\left|{\boldsymbol \xi}\right|\chi_{e}\right\rangle\right|^{2}\\
  \label{eq: Gammaaug}
  \Gamma_{\text{aug}}({\bf R})&=\int \frac{d^{3}k'}{(2\pi)^{2}}|\mathcal{V}^{AB}({\bf k'},{\bf R})|^{2}\delta(\varepsilon_{k}'+\epsilon_{g}-\varepsilon_{g}-\epsilon_{e}).
  \end{align}  
  \subsection{Ratio of cross sections at fixed ${\bf R}$}
In order to compare the two-center photoionization, we also calculate the direct photoionization of atom $A$. The corresponding transition amplitude is given by
\begin{equation}
\label{eq: S1}
S^{(1)}=-i\int\limits_{-\infty}^{\infty}dt\,\langle \varphi_{{\bf k}}({\bm r})|\hat{W}_{A}(\omega)| \varphi_{g}({\bm r})\rangle e^{-i(\varepsilon_{g}+\omega-\varepsilon_{k})t}.
\end{equation}
Here, the operator $\hat{W}_{A}=\frac{{\bf A_{0}\cdot\hat{p}}_{r}}{2c}$ describes the coupling of the electron in atom $A$ to the electromagnetic field.
The direct photoionization process can interfere with the indirect process leading to the same final state. The resulting transition amplitude thus consists of two individual amplitudes
\begin{eqnarray}
 S^{(12)}=S^{(1)}+S^{(2)}.
 \end{eqnarray} 
We then obtain the transition cross sections $\sigma^{(1)}$, $\sigma^{(2)}$ and $\sigma^{(12)}$ for the one-center, two-center and total photoionization processes, respectively; by virtue of
\begin{eqnarray}
\label{eq: sigmaallg}
\sigma^{(M)}=\frac{1}{\tau j}\int\frac{d^{3}k}{(2\pi)^{3}}|S^{(M)}|^{2},
\end{eqnarray}
where $\tau$ denotes the interaction time and $j=\frac{\omega A_{0}^{2}}{8\pi c}$ is the incident photon flux.
The cross sections for two-center and direct one-center photoionization can be related. Fixing ${\bf R}=R{\bf e}_{z}$, one obtains:
\begin{align}
\sigma^{(2)}&=\frac{1}{j}\frac{4}{R^{6}}\int \frac{d^{3}k}{(2\pi)^{3}}\left|\left\langle\varphi_{{\bf k}}\chi_{g}\left|z \xi_{z}\right|\varphi_{g}\chi_{e}\right\rangle\right|^{2}\times\nonumber\\
& 2\pi\delta(E_{{\bf k},g}-E_{g,g}-\omega)\frac{\left|\left\langle\chi_{e}\left|\frac{1}{2c}{\bf\hat{A}\cdot\hat{p}}_{\xi}\right|\chi_{g}\right\rangle\right|^{2}}{\Delta^{2}+\frac{1}{4}\Gamma^{2}}\nonumber\\
&\approx \sigma^{(1)}\frac{\Gamma_{\text{rad}}^{2}}{\Delta^{2}+\frac{1}{4}\Gamma^{2}}\left(\frac{3c^{3}}{4R^{3}\omega^{3}}\right)^{2},
\end{align}
On resonance, the ratio reads:
\begin{equation}
\label{eq: Abschaetzung_atom}
\frac{\sigma^{(2)}}{\sigma^{(1)}}\approx \frac{\Gamma_{\text{rad}}^{2}}{\Gamma^{2}}\left(\frac{3c^{3}}{2R^{3}\omega^{3}}\right)^{2}.
\end{equation}
$\Gamma_{\text{rad}}$ and $\Gamma_{\text{aug}}(R)$ can be calculated from Eqs. \eqref{eq: Gamma} and \eqref{eq: Gammaaug} or obtained from literature.
\subsection{Inclusion of nuclear motion}
\subsubsection{Energy shift}
\label{energy shift}
\begin{figure}
\begin{center}
\includegraphics[width=0.47\textwidth]{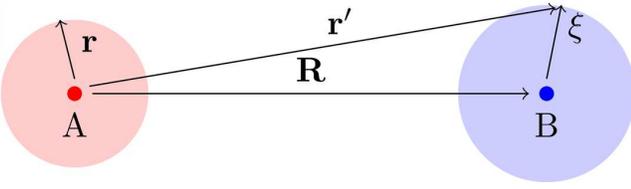}
\end{center}
\caption{Scheme of the spatial dependencies considered in the calculation of the energy shift.}
\label{figure2}
\end{figure}

The presence of an atom in the proximity of another atomic center leads to an appearance of an interaction between them, described by a static potential. As a consequence, the energy of the system depends on their separation ${\bf R}$.
We therefore include interactions of Coulombic and exchanging nature as first order perturbations. Also van-der-Waals terms have to be included in the picture as they generally become more relevant than first order shifts for larger distances \cite{Patil}. 
The Coulomb correction to the energy is given by
\begin{eqnarray}
V_{\text{coul}}(R)&=&\int d^{3}r d^{3}\xi \Big(\frac{\beta_{1}}{R}-\frac{\beta_{2}}{|{\bf R}-{\bf r}|}-\frac{\beta_{3}}{|{\bf R}+{\boldsymbol \xi}|}\nonumber\\
&&+\frac{\beta_{4}}{|{\bf R}-{\bf r}+{\boldsymbol \xi}|}\Big)|\Phi({\bf r},{\boldsymbol \xi})|^{2},
\end{eqnarray}
where the spatial dependencies are shown in Fig. \ref{figure2}.
Here, $\beta_{m}$ are coefficients containing the charges of particles involved in the respective interaction.\\ 
The exchange energy takes the indistinguishability of the electrons into account and is given by
\begin{align}
V_{\text{exc}}(R)&=-\frac{1}{2}\int d^{3}r d^{3}\xi \Big(\frac{\beta_{1}}{R}-\frac{\beta_{2}}{|{\bf R}-{\bf r}|}-\frac{\beta_{3}}{|{\bf R}+{\boldsymbol \xi}|}\nonumber\\
&+\frac{\beta_{4}}{|{\bf R}-{\bf r}+{\boldsymbol \xi}|}\Big)\Phi({\bf r},{\bf R}+{\boldsymbol \xi})\Phi^{*}({\bf R}+{\boldsymbol \xi},{\bf r}).
\end{align}
 The factor of $\frac{1}{2}$ is due to the spin. In general, the van-der-Waals interaction can be calculated as a second order correction of the form
\begin{eqnarray}
V_{\text{vdW}}(R)=\sum_{n'\neq m}\frac{|\langle \Phi_{m}|\hat{V}_{AB}|\Phi_{n'}\rangle |^{2}}{E_{n'}^{(0)}-E_{m}^{(0)}}.
\end{eqnarray}
However, when possible, we use literature values for better accuracy , where higher order terms ($\frac{C_{8}}{R^{8}}$,$\frac{C_{10}}{R^{10}}$) can be included. Consequently, the van-der-Waals interaction can be described by
\begin{equation}
V_{\text{vdW}}(R)=-f_{6}(R)\frac{C_{6}}{R^{6}}-f_{8}(R)\frac{C_{8}}{R^{8}}-f_{10}(R)\frac{C_{10}}{R^{10}}.
\end{equation}
Literature values require the restriction to one orientation of the internuclear vector ${\bf R}$.
The Van-der-Waals coefficients are multiplied by damping factors \cite{Patil} 
\begin{equation}
 f_{2L+4}(R)=1-e^{-R/\overline{a}}\sum_{n=0}^{2L+7}\frac{1}{n\!}\left(\frac{R}{\overline{a}}\right)^{n}.
 \end{equation} 
Here, $\overline{a}$ is defined by the ionization energies of both constituents $A$ and $B$:
 $\overline{a}=\frac{1}{4}\left(\frac{1}{\sqrt{2 \varepsilon_{g}}}+\frac{1}{\sqrt{2 \epsilon_{g}}}\right)$.
The damping functions are introduced to include the effects of charge overlap \cite{damping}.

The total static interaction energy between the atoms is then given by the sum of these three interactions
\begin{equation}
V_{\text{int}}({\bf R})=V_{\text{coul}}+V_{\text{exc}}+V_{\text{vdW}}.
\end{equation}
\subsubsection{Vibrational States}
\label{sec:vibrational states}

The interaction of neighbouring atoms $A$ and $B$ can lead to a complex of these two constituents which cannot be treated individually anymore.
 This bound system is then characterized not only by its motion as a whole but also by its relative motion. The rotation and vibration of diatomic molecules  influences the binding energy on different energy scales. Since the rotational motion creates levels with a spacing much smaller than the levels due to vibrational motion, we will restrict our calculations to vibrational effects solely \cite{Bransden}.
We consider vibrational levels and wave functions for the three electronic states described in section \ref{sec:general}.

(I) $\Psi_{g,g}=\psi_{\text{i}}(R,\nu_{\text{i}})$ refers to the vibrational wave function for the  intial system including both atoms $A$ and $B$ in their electronic ground states. The corresponding vibrational energy reads $E_{\text{i}}(\nu_{\text{i}})=-D_{\text{i}}+E_{\text{vib}}(\nu_{\text{i}})$, where $D_{\text{i}}$ is the maximum depth of the corresponding interaction potential curve.
(II) $\Psi_{g,e}=\psi_{\text{a}}(R,\nu_{\text{a}})$ denotes the vibrational wave function for the intermediate system of an autoionizing state with $A$ in its electronic ground state and $B$ excited via photoexcitation. The vibration leads to the energy shift $E_{\text{a}}(\nu_{\text{a}})=-D_{\text{a}}+E_{\text{vib}}(\nu_{\text{a}})$.
(III) $\Psi_{{\bf k},g}=\psi_{\text{f}}(R,\nu_{\text{f}})$ describes the final  vibrational wave function  with $A$ being ionized and $B$ back in its electronic ground state. The vibration leads to the energy shift $E_{\rm{f}}(\nu_{\rm{f}})=-D_{\rm{f}}+E_{\rm{vib}}(\nu_{\rm{f}})$.\\
We calculate the vibrational wave functions and the energy levels from the potential energy shift. 
In order to avoid a numerical computation of the vibrational energies and states, we fit a Morse potential to our potential curves in order to find the necessary parameters. The potential reads
\begin{eqnarray}
\label{eq:Morse}
V_{\rm{Morse}}(R)=D\left[\left(1-e^{-\alpha (R-R_{\rm{eq}})}\right)^{2}-1\right],
\end{eqnarray}
where $D$ is the depth of the potential, $R_{\text{eq}}$ is the equillibrium distance and $\alpha$ describes the width of the potential.
Employing this fit, we obtain the vibrational energy levels 
\begin{equation}
\label{eq:Morse_Energie}
E_{\rm{vib}}(\nu)=\sqrt{\frac{2D\alpha^{2}}{\mu}}\left[\left(\nu+\frac{1}{2}\right)-\frac{1}{\kappa}\left(\nu+\frac{1}{2}\right)^{2}\right]
 \end{equation} 
 with $\kappa=\sqrt{\frac{8\mu D}{\alpha^{2}}}$ and the reduced mass of the system $\mu$. 
The parameters obtained by the fit are then inserted in the vibrational wave function \cite{Gallas}:
\begin{eqnarray}
\psi(R,\nu)=\sqrt{\frac{\alpha b (\nu!)}{\Gamma(\kappa-\nu)}}e^{-z/2}z^{(\kappa-2\nu-1)/2}\mathcal{L}_{\nu}^{(\kappa-2\nu-1)}(z)
\end{eqnarray}
with the associated Laguerre polynomials $\mathcal{L}_{\nu}^{(\kappa-2\nu-1)}(z)$, $b=\kappa-2\nu-1$ and $z=\kappa e^{-\alpha (R-R_{\rm{eq}})}$. 

The vibrational wave functions contribute to the cross section by the insertion of Franck-Condon factors, which describe the overlap of the vibrational nuclear wave functions  associated with the two electronic states involved in a transition. The factor therefore favours transitions with large overlaps, involving no substantial changes of the nuclear motion in terms of position and kinetic energy \cite{LaForge}. Photoionization, photoexcitation and the energy transfer via dipole-dipole interaction do not limit the possible combinations of vibrational states. Every matrix element describing an electronic transition contains an integration over $R$ as well. The general form of a transition matrix element is given by:
	    \begin{equation}
	 \mathcal{M}=   \langle \Phi_{2}^{el}({\bf r},{\boldsymbol \xi})\Psi_{2}(R,\nu_{2})|\hat{V}({\bf R},{\bf r},{\boldsymbol \xi})|\Phi_{1}^{el}({\bf r},{\boldsymbol \xi})\Psi_{1}(R,\nu_{1})\rangle,
	    \end{equation}
	     where $\hat{V}$ is an arbitrary operator, $i$ and $f$ refer to the inital and final states of the transition. The matrix element depends on the vibrational levels $\nu$.
	     
We do not apply the Condon approximation in our calculation since $\hat{V}_{AB}$ depends heavily on $R$. Therefore, the corresponding Franck-Condon factor cannot be separated from the matrix element consisting of electronic and vibrational transitions.
The vibrational levels $\nu$ for the intermediate and final state  give rise to a multitude of transitions. Hence, we have to sum coherently over all possible intermediate states in the transition amplitude $S^{(2)}_{\text{mol}}$ and incoherently over all possible final states within the associated two-center cross section. 

Considering the vibrational wave functions, the previously position-dependent  $\Gamma_{\text{aug}}(R)$ is transformed into $\bar{\Gamma}_{\text{aug}}$. For this purpose we average $\Gamma_{\text{aug}}(R)$ over the probability density $\left|\psi_{a}(R,\nu_{a})\right|^{2}$ of the intermediate vibrational state for every vibrational level $\nu_{a}$ \cite{Fussnote3}. For the radiative width the atomic value from Eq. \eqref{eq: Gamma} is inserted because $\Gamma_{\text{rad}}$ remains practically unaltered in the presence of the neighbour atom.

Including vibrational levels, vibrational wave functions and decay rates, the explicit expressions accounting for the molecular effects read
\begin{widetext}
\begin{align}
S^{(2)}_{\text{mol}}=-2\pi i\sum_{\nu_{\text{a}}}\delta(E_{{\bf k},g}+E_{\text{f}}(\nu_{\text{f}})-E_{g,g}-E_{\text{i}}(\nu_{\text{i}})-\omega)\frac{\langle \Phi_{{\bf k},g}\Psi_{{\bf k},g}|\hat{V}_{AB}({\bf R})|\Phi_{g,e}\Psi_{g,e}\rangle \langle\Phi_{g,e}\Psi_{g,e}|\hat{W}_{B}(\omega)|\Phi_{g,g}\Psi_{g,g}\rangle}{\epsilon_{g}+E_{\text{i}}(\nu_{\text{i}})+\omega-\epsilon_{e}-E_{\text{a}}(\nu_{\text{a}})+\frac{i}{2}(\Gamma_{\text{rad}}+\bar{\Gamma}_{\text{aug}})}
\end{align}
\begin{align}
S^{(1)}_{\text{mol}}=-2\pi i \delta(E_{{\bf k},g}+E_{\text{f}}(\nu_{\text{f}})-E_{g,g}-E_{\text{i}}(\nu_{\text{i}})-\omega) \langle\Phi_{{\bf k},g}\Psi_{{\bf k},g}|\hat{W}_{A}(\omega)|\Phi_{g,g}\Psi_{g,g}\rangle
\end{align}
\end{widetext}
 Since $\hat{W}_{A}$ and $\hat{W}_{B}$ do not depend on $R$, the corresponding Franck-Condon factors can be separated from the electronic transition matrix elements. Therefore, 
\begin{align}
F_{\text{i},\text{a}}(\nu_{\text{i}},\nu_{\text{a}})&=\int dR \, \psi_{\text{i}}(R,\nu_{\text{i}})\psi^{*}_{\text{a}}(R,\nu_{\text{a}})\\
F_{\text{i},\text{f}}(\nu_{\text{i}},\nu_{\text{f}})&=\int dR \, \psi_{\text{i}}(R,\nu_{\text{i}})\psi^{*}_{\text{f}}(R,\nu_{\text{f}}).
\end{align}
For $F_{\text{a},\text{f}}(\nu_{a},\nu_{f})$ however, one has to incorporate the $R$-dependency of $\hat{V}_{AB}$. The factor in question can be simplified to $1/R^{3}$, see Eq. \eqref{eq: VAB}. This yields
\begin{equation}
F_{\text{a},\text{f}}(\nu_{\text{a}},\nu_{\text{f}})=\int dR\,\frac{ \psi_{\text{a}}(R,\nu_{\text{a}})\psi^{*}_{\text{f}}(R,\nu_{\text{f}})}{R^{3}}.\\
\end{equation}
\section{Results and Discussion}
\label{Results}
Let us apply the theory established in Sec. \ref{sec:theory} to the photoionization of a weakly bound van-der-Waals molecule, which is chosen as a $^{7}$Li$^{4}$He dimer. The internuclear axis is taken along the $z$-axis, which also serves as the quantization axis. 
We choose effective electronic states $\chi$ and $\varphi$ in order to achieve a reasonable comparableness to the atomic species under consideration. 
Neither Li nor He are single-electron atoms. To describe He as a two-electron system we employ symmetrized 
superpositions of product states formed by hydrogen-like wave functions 
with an effective charge $Z_{\rm He}$. The He ground state 
correspondingly reads $\chi_{1s}({\boldsymbol \xi}_{1}) \chi_{1s}({\boldsymbol \xi}_2)$, 
and the excited state is given by $\frac{1}{\sqrt{2}}[\chi_{1s}({\boldsymbol \xi}_{1}) 
\chi_{2p_{0}}({\boldsymbol \xi}_{2})+\chi_{2p_{0}}({\boldsymbol \xi}_{1}) \chi_{1s}({\boldsymbol \xi}_{2})]$. Within such an approach one can calculate the matrix elements including helium by using only a one-particle wave function $\chi({\bf \xi})$ and then multiplying by  a factor of 2 contributing for the superposition. Note, that this also applies to the two-center Auger decay width $\Gamma_{\text{aug}}$ for both the atomic and molecular consideration. 
For the valence electron of the alkali atom lithium in its ground state, we choose the Bates-Darngaard wave function as in \cite{Patil}, so that
\begin{eqnarray}
\varphi_{g}({\bf r})=\frac{1}{\sqrt{8\pi}}\frac{1}{\Gamma(a+1)}\left(\frac{2}{a}\right)^{a+\frac{1}{2}}r^{a-1}\left(1+\frac{v}{r}\right)e^{-r/a}
\end{eqnarray}
 with $a=\frac{1}{\sqrt{2|\varepsilon_{g}|}}$ depending on the binding energy and $v=-\frac{1}{2}a^{2}(a-1)$. Note that $a$ corresponds to an effective nuclear charge $Z_{\text{Li}}\approx 1.259$ for $\varepsilon_{g}= 5.39\text{eV}$ \cite{NIST}.

After ionization, the influence of the remaining lithium ion on the emitted electron of Li is accounted for by employing a Coulomb wave $\varphi_{{\bf k}}$ \cite{LandauQM}. For this state, the same effective nuclear charge $Z_{\text{Li}}$ as in the ground state is chosen.
In accordance with Eq. \eqref{eq: sigmaallg}, this continuum state is normalized to a quantization volume of unity.\\
Within this system, we consider the dipole-allowed transition $(1s-2p_{0})$ for the photo-excitation of atom $B$. The transition energy defines the effective nuclear charge for both states. With an excitation energy $\omega_{\text{He}}=\text{21.218 \text{eV}}$ \cite{NIST}, we find $Z_{\text{He}}=1.435$ . 
We have now captured some basic features of the Li-He dimer.

\subsubsection{Results at fixed nuclei}
Before incorporating molecular effects into our description of the system  \cite{Friedrich}, let us first calculate the cross sections for the ionization of lithium by disregarding the molecular structure and rather considering Li and He as two unbound atoms which are separated by a fixed internuclear distance $R$ on which the two-center cross section strongly depends \cite{2CPI1,2CPI2}.
 
\begin{figure}[t]
\includegraphics[width=0.49\textwidth]{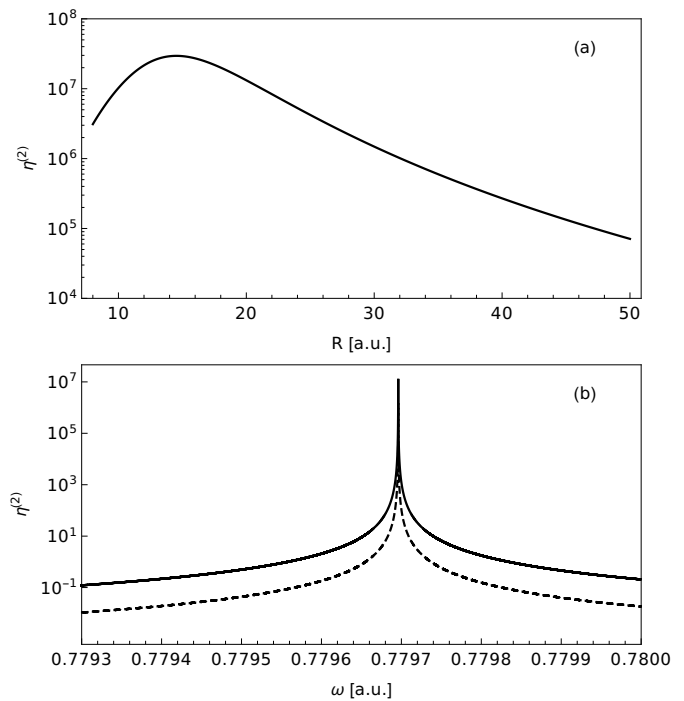}
\caption{Ratios $\eta^{(2)}_{\text{atomic}}(\omega,R)=\frac{\sigma^{(2)}_{\text{atomic}}}{\sigma^{(1)}_{\text{atomic}}}$ in a Li-He system, ignoring molecular effects. (a) $\eta^{(2)}_{\text{atomic}}(\omega,R)$ plotted against the internuclear distance $R$ on resonance. (b) $\eta^{(2)}_{\text{atomic}}(\omega,R)$ plotted against the incident photon energy $\omega$ for fixed internuclear distance $R=20a_{0}$ (solid) and $R=30a_{0}$ (dashed). } 
\label{fig:ohne_Kernbewegung}
\end{figure}

Due to the atomic approach, the two-center Auger decay width $\Gamma_{\text{aug}}$ in Eq. \eqref{eq: Gammaaug} is very sensitive to $R$. Both the decay width $\Gamma_{\text{aug}}$ and the square of the matrix element of the radiationless energy transfer include the dependency on $\frac{1}{R^{6}}$. On resonance, the $R$-dependence of the ratio $\eta_{\text{atomic}}^{(2)}(\omega,R)=\sigma^{(2)}_{\text{atomic}}/\sigma^{(1)}_{\text{atomic}}$ is depicted in Fig. \ref{fig:ohne_Kernbewegung} (a). For relatively small distances $R$, the ratio of cross sections shows a  steep incline for increasing  spacings. For sufficiently large separations, $\Gamma_{\text{aug}}$ becomes small in comparison to the constant radiative decay width  $\Gamma_{\text{rad}}$. Consequently, the saturation of the total decay width $\Gamma$ sets in quickly. The dependence on $R$ of the matrix element mentioned above, however, does not exhibit saturation. Therefore, the cross section ratio depicted in Fig.\ref{fig:ohne_Kernbewegung} (a) shows the mentioned $\frac{1}{R^{6}}$ behaviour after the maximum is reached. 

For fixed internuclear separations $R=20\,a_{0}$ and $R=30\,a_{0}$, the ratio of cross sections is plotted against the incident photon energy $\omega$ in Fig. \ref{fig:ohne_Kernbewegung} (b). The plot shows a single peak where the photon energy is resonant with the electronic $1s$-$2p$ transition in He. The peak height demonstrates that 2CPI can strongly dominate over the direct photoionization.
\subsubsection{Effects of nuclear motion}
Until now, we have not taken any effects of the molecular bond in Li-He into account.
In neighbourhood of each other, the influence of the two atoms leads to a potential surface creating a possibly bound state.
We now consider these potential energy curves for the three electronic states participating in the process depending on the internuclear distance $R$.

Note that for the initial state, where both atoms are in their ground state, the direction of $R$ does not play a role. For the intermediate state, however, where helium is excited, the interaction depends on the direction of ${\bf R}$. Here, ${\bf R}=R{\bf e}_{z}$ is considered, as mentioned above.

For the chosen geometry, the van-der-Waals coefficients can be given for inital, intermediate and final electronic state of Li-He.
(I): For the initial state, we find $C_{6}=22.5 \text{ a.u.}$, $C_{8}=1.06\times10^{3}\text{ a.u.}$ \cite{Patil}.
(II): The intermediate state is described using, $C_{6}=6123\text{ a.u.}$, $C_{8}=7.85\times10^{5}\text{ a.u.}$, $C_{10}=1.02\times10^{8}\text{ a.u.}$ \cite{vdW_2p}.
(III): For the ionic system of the final state, $C_{6}=0.298\text{ a.u.}$, $C_{8}=1.98\text{ a.u.}$ are employed. This way, we obtain the relevant terms as calculated in \cite{Soldan}.

\begin{figure}
\includegraphics[width=0.49\textwidth]{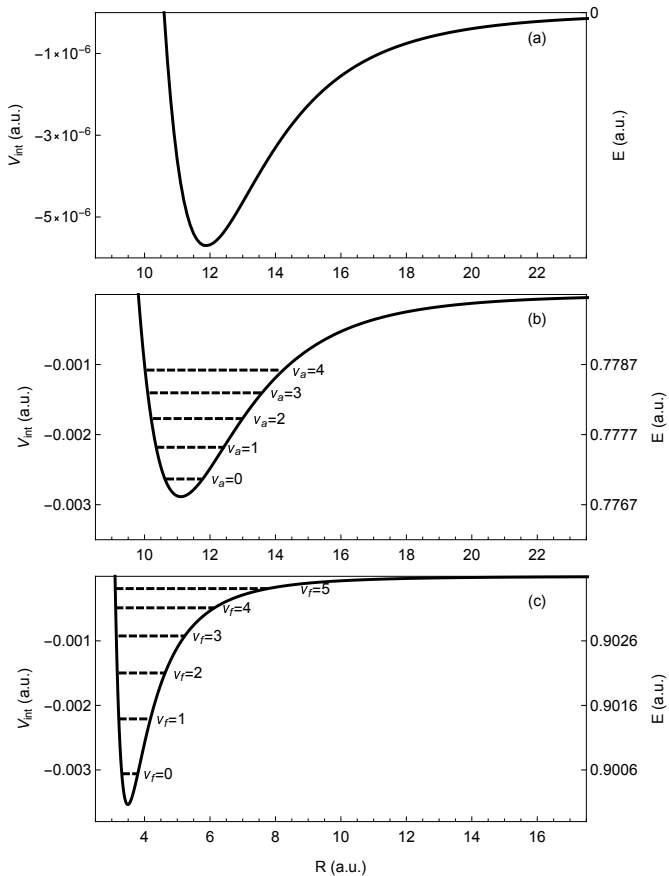}
\caption{Potential energy curves (solid) of the electronic states involved in the two-center photoionization process of Li-He and some corresponding vibrational levels (dashed). (a) Initial state.  The only supported vibrational level $\nu_{\text{\text{i}}}=0$ exhibits a binding energy  that is too small to be graphically resolved here. (b) Intermediate state with the lowest-lying vibrational levels. 
In total, 11 vibrational levels ($0\le\nu_a\le 10$) associated with bound 
molecular states were obtained; all of them are included in our numerical 
calculations.
(c) Final state including 6  bound vibrational levels ($0\le\nu_{\text{f}}\le 5$). As an absolute scale, the potential energy $E$ relative to the ground state energy at infinite internuclear distance is depicted on the right side. }
\label{fig:energyshift}
\end{figure}

In Fig. \ref{fig:energyshift}, the interaction potential curves of the relevant electronic states of Li-He are depicted. For the initial state $\Phi_{g,g}$ and the intermediate state $\Phi_{g,e}$ the potential minima are close to each other. The final state $\Phi_{{\bf k},g}$, however, features its minimum at a much smaller distance. The depths of the potential for the final and intermediate state are comparable in magnitude.

The initial state, however, exhibits a much more shallow potential curve.
This leads to the initial state supporting only one vibrational level. The binding energy of merely $6\text{mK}=1.9\times 10^{-8}\text{a.u.}$ \cite{Friedrich} is not depicted graphically here since it is too close to the plot axis. The vibrational level expands widely and 
spans over a large range of interatomic distances $R$ allowing for a very large bond length. In the results from  \cite{Friedrich}, the equillibrium distance $R_{\text{eq}}\approx 11.3 a_{0}$, representing the minimum of the potential, is much smaller than the mean distance $R \approx 55a_{0}$.
Literature values for the potential curves can be found for the initial and final state \cite{Patil, Soldan}and show good agreement with our calculations (within about $10\%$). For the intermediate states, no literature values beyond the van-der-Waals coefficients were available.

From the potential curves depicted in Fig.\ref{fig:energyshift}, the parameters for a Morse potential [see Eq. \eqref{eq:Morse} and Table \ref{tab: Morse}] are extracted.
\begin{table}[h]
\begin{center}
\begin{tabular}{|c||c|c|c| }
&ground state & intermediate state & final state\\
 \hline
$R_{\text{eq}}$&11.9&11.0&3.5\\
$D$& $5.7\times10^{-6}$&$2.9\times 10^{-3}$&$3.5\times 10^{-3}$ \\ 
$\alpha$&0.43&0.44&0.80
\end{tabular}
 \caption{Fitted Morse parameters (in a.u.) for the interaction potential curves. }\label{tab: Morse}
\end{center}\end{table}
 The resulting vibrational energy levels are calculated  according to Eq. \eqref{eq:Morse_Energie} and also shown in the figure. The inclusion of molecular effects leads to a shift concerning the energy of the Li-He system  for every electronic state. Compared to the atomic consideration, the resonance energies of the system are blueshifted \cite{Hergenhahn}.

The vibrational wave functions are calculated from the obtained Morse potential as described in Sec. \ref{sec:vibrational states}.
Since the positions of the potential minima for the initial and intermediate state are quite close to each other, their corresponding vibrational wave functions yield a significant overlap for all intermediate vibrational levels $\nu_{\text{\text{a}}}$. The potential minimum of the final state, however, is strongly shifted to smaller interatomic distances $R$. An overlap with the other sets of states $\lbrace \psi_{\text{a}}(R,\nu_{\text{a}})\rbrace$, $\lbrace \psi_{\text{i}}(R,\nu_{\text{i}})\rbrace$ is only present for high vibrational levels $\nu_{\text{\text{f}}}$. This results in a strong dependence of the Franck-Condon factors on the combination of vibrational levels.

For the calculation including molecular effects, the decay widths have to be evaluated. Using the wave functions mentioned in Sec. \ref{sec:general}, $\Gamma_{\text{rad}}^{\text{He}}=6.62\times10^{-8}\,a.u.$
 is obtained. This value differs from the literature value $\Gamma_{\text{rad}}^{\text{He}}=4.35\times 10^{-8}\,a.u.$ \cite{NIST} by a factor of about $1.5$. However, since the wave functions are inserted in all matrix elements of both processes, we use our value for selfconsistency. The two-center Auger decay rate $\Gamma_{\text{aug}}(R)$  in Eq. \eqref{eq: Gammaaug} depends on the internuclear distance $R$ and therefore is influenced by the nuclear motion. The two-center Auger decay is only allowed for distances greater than $\approx 4.3a_{0}$, where our potential curves for intermediate and final states cross \cite{Scheit}.

Before showing our numerical results, the analytical approach provides the possibility to obtain an approximate formula displaying the influence of the nuclear motion on the ratio of cross sections $\eta^{(2)}(\omega)=\frac{\sigma^{(2)}(\omega)}{\sigma^{(1)}(\omega)}$. With the restriction to one combination of vibrational levels $\nu_{\text{a}}$ and $\nu_{\text{f}}$, an approximation of $\eta^{(2)}$ can be given:
\begin{equation}
\label{eq: FC_Vereinfachung}
\frac{\sigma^{(2)}}{\sigma^{(1)}}\Bigg|_{\omega=\omega_{\text{res}}}\approx \frac{9c^{6}}{\omega_{\text{res}}^{6}R_{\text{eq}}^{6}}\frac{\Gamma_{\text{rad}}^{2}}{\Gamma^{2}}\left(\frac{R_{\text{eq}}^{6}\left(F_{\text{i,a}}F_{\text{a,f}}\right)^{2}}{\left(F_{\text{i,f}}\right)^{2}}\right),
\end{equation}
where $\Gamma$ already includes the two-center Auger width $\bar{\Gamma}_{\text{aug}}$ incorporating the nuclear motion. Note, that the resonant energy  $\omega_{\text{res}}$ depends on the vibrational energy levels involved. Ignoring molecular effects, $\omega_{\text{res}}=\omega_{B}$ meets the criterion for  resonance. 

A comparison of Eq. \eqref{eq: FC_Vereinfachung} with Eq. \eqref{eq: Abschaetzung_atom}, which was obtained in an atomic picture, reveals the influence of molecular effects on 2CPI. Considering the photoionization processes including molecular effects, the   formula in Eq. \eqref{eq: Abschaetzung_atom} is modified by multiplying the term in parentheses including the Franck-Condon overlaps. $R_{\text{eq}}$ has been inserted in the parentheses of Eq. \eqref{eq: FC_Vereinfachung} to make the ratio of Franck-Condon factors dimensionless, since $F_{\text{a,f}}$ includes the $R$-dependence of the electron-electron interaction. Note, that the ratios  for Li-He calculated from an atomic point of view are still dependent on $R$. Besides, the expression in Eq. \eqref{eq: Abschaetzung_atom} differs by a factor of $4$ due to the fact that here, we consider helium as a two-electron system.

\begin{figure}
\includegraphics[width=0.49\textwidth]{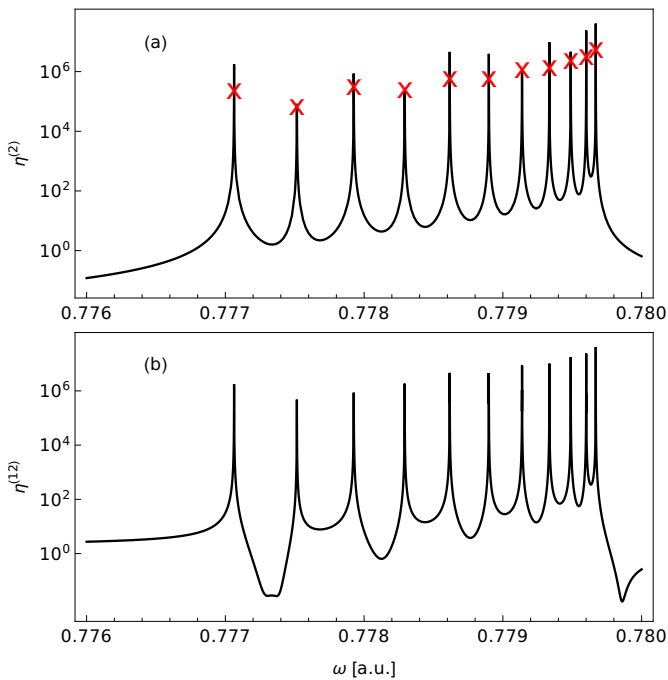}\\
\caption{Ratios of cross sections in a Li-He dimer including the vibrational levels $\nu_{\text{a}}\in\left\lbrace 0,1,2,..,10\right\rbrace $ and $\nu_{\text{f}}\in\left\lbrace 0,1,2,3,4,5\right\rbrace $. a) $\eta^{(2)}(\omega)$ and values from the approximate expression in Eq \eqref{eq: FC_Vereinfachung} (crosses). b) $\eta^{(12)}(\omega)$ containing interference of direct and two-center photoionization} 
\label{fig:P2/P1}
\end{figure}

It should be mentioned that the full calculation of the 2CPI cross section including vibrational effects calls for the summation over $\nu_{\text{a}}$ of the transition amplitudes as well as for the summation over $\nu_{\text{f}}$ in the cross section. Therefore, Eq. \eqref{eq: FC_Vereinfachung} can only indicate estimated values for particular vibrational transitions. As we will see below, for the strongest transitions $\nu_{\rm{f}}=5$ the values computed from Eq. \eqref{eq: FC_Vereinfachung} are only slightly reduced as compared with $\eta_{\text{atomic}}^{(2)}(\omega_{B},R)$. The approximated values are depicted by crosses for individual $\nu_{\rm{a}}$ in Fig. \ref{fig:P2/P1} a).
 
With these preparations we now proceed to the numerical computation of the ratios $\eta^{(2)}(\omega)=\frac{\sigma^{(2)}(\omega)}{\sigma^{(1)}(\omega)}$ and $\eta^{(12)}(\omega)=\frac{\sigma^{(12)}(\omega)}{\sigma^{(1)}(\omega)}$. We include the vibrational levels $\nu_{\text{a}}=0,1,...,10$ and $\nu_{\text{f}}=0,1,..,5$.
While the direct photoionization process exhibits a cross section smoothly depending on the incident photon energy $\omega$, the two-center process is a  resonant one \cite{2CPI1,2CPI2,2CPIexp}. Including the various combinations of vibrational levels, the single atomic resonance peak  of Fig. \ref{fig:ohne_Kernbewegung} b) is expected to fan out with respect to the photon energy $\omega$.
The ratios depicted in Fig. \ref{fig:P2/P1} show a multiplet of peaks caused by the various vibrational levels of the intermediate molecular state. They are responsible for the shift of the particular resonant energy. The vibrational level of the final state does not affect the resonance condition and therefore causes no additional splitting. The peak on the most lefthand side can be associated with  vibrational level $\nu_{\text{a}}=0$ and $\nu_{\text{a}}$ is ascending from left to right for the other peaks.

The ratio $\eta^{(12)}$ considering the total cross section in Fig. \ref{fig:P2/P1} b) shows a sequence of Fano profiles (alternating  in phase between adjacent vibrational levels $\nu_{\text{a}}$ and $\nu_{\text{a}}+1$). The cross section $\sigma^{(12)}$ includes both the direct and indirect photoionization processes which allows for interference and therefore leads to the Fano profiles including minima and maxima.
However, the Fano profile of a given resonance is strongly modified due to the summation over the final vibrational level $\nu_{\text{f}}$. The characteristics of the minimum differ for each level $\nu_{\text{f}}$. Considering $\eta^{(12)}$ for each $\nu_{\rm{f}}$ individually, the peak position remains the same while the position and depth of the minimum changes. As a result, some Fano minima are cancelled by the overlap of all $\nu_{\text{f}}$. A relatively deep minimum can only be seen in between $\nu_{\text{a}}=0$ and $\nu_{\text{a}}=1$ and next to $\nu_{\text{f}}=10$, where the individual components overlap in order to form a minimum. 
This modification represents a strong contrast to the atomic calculation \cite{2CPI1}, where the Fano profile is much more pronounced.
Off resonance, the ratio eventually falls off to $1$. The ratio $\eta^{(12)}$ shown in Fig. 5b) involves incoherent summations
over the final vibrational states, separately in the numerator and the
denominator. In order to obtain a pure Fano profile, a fixed value of
$\nu_{\text{f}}$ must be considered instead. For $\nu_{\text{f}}=5$, representing the strongest
contribution to a peak, the Fano parameters q are given in Table \ref{tab:Fano}.
While the sign of the Fano parameters alternates, their absolute value,
roughly speaking, increases from small to large $nu_{\text{a}}$. Being on the order
of $10^{3}$, they resemble the corresponding values obtained for fixed nuclei
(see caption of Table \ref{tab:Fano}).

\begin{table}[h]
\begin{center}
\begin{tabular}{|c|c||c|c||c|c| }
$\nu_{\text{a}}$& $q$&$\nu_{\text{a}}$&$q$&$\nu_{\text{a}}$& $q$\\
 \hline
0& $1.0\times 10^{3}$&4 & $1.7\times 10^{3}$&8 & $3.3\times 10^{3}$    \\  
1 & $-5.6\times 10^{2}$ &5 & $-1.7\times 10^{3}$ & 9 & $-3.8\times 10^{3}$   \\   
2 & $1.2\times 10^{3}$ &6 & $2.3\times 10^{3}$   &10 & $5.0\times 10^{3}$  \\  
3 & $-1.1\times 10^{3}$  &7 & $-2.5\times 10^{3}$&&  \\ 
\end{tabular}
 \caption{Approximated Fano parameters $q$ for each vibrational state $\nu_{\text{a}}$ for $\nu_{\text{f}}=5$. For fixed nuclei and $R=10 \text{a.u.}$ ($R=20 \text{a.u.}$), one finds $q \approx 3.2\times 10^{3}$ ($q \approx 3.6\times 10^{3}$).}\label{tab:Fano}
\end{center}\end{table}

\begin{figure}[h]
\includegraphics[width=0.49\textwidth]{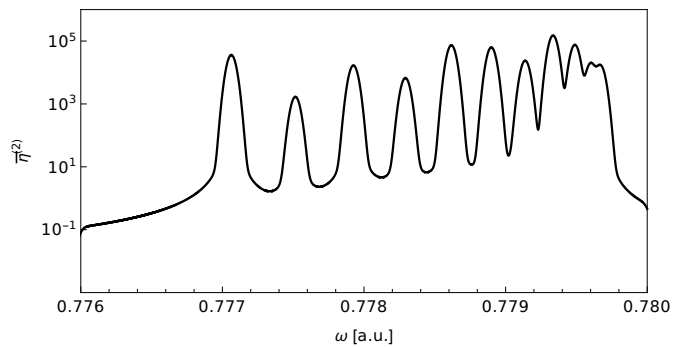}
\caption{Ratio  $\bar{\eta}^{(2)}=\frac{\bar{\sigma}^{(2)}}{\bar{\sigma}^{(1)}}$ for a Li-He dimer.}
\label{fig:Mittelung}
\end{figure}

The orders of magnitude of the peak heights shown  in Fig. \ref{fig:P2/P1}  a) are comparable to those in Fig \ref{fig:ohne_Kernbewegung} b) and Ref. \cite{2CPI1}. Therefore, the inclusion of molecular effects does not change the magnitude of the resonant enhancement significantly. We note that from left to right, i.e. from
$\nu_{\text{a}}=0$ to $\nu_{\text{a}}=10$, the peaks become more narrow
and have a tendency to reach larger maximum values.

In order to further compare the calculations for fixed nuclei with the calculations involving molecular effects, the total decay widths of the peaks can be considered in table \ref{tab: 1}. They have been obtained by analyzing the data from the numerical calculations. We see that the total width decreases with the increase of the vibrational level of the intermediate state $\nu_{\text{a}}$. This is because the internuclear distance tends to grow for higher lying vibrational excitations, as illustrated in Fig. \ref{fig:energyshift}. The same trend of decreasing ICD widths has also been observed in He-Ne dimers \cite{2CPIexp,Mhamdi18}.   
\begin{table}[h]
\begin{center}
\begin{tabular}{|c|c||c|c||c|c| }
$\nu_{\text{a}}$& \textbf{$\Gamma$}(a.u.)&$\nu_{\text{a}}$& \textbf{$\Gamma$}(a.u.)&$\nu_{\text{a}}$& \textbf{$\Gamma$}(a.u.)\\
 \hline
0& $5.3\times 10^{-7}$&4 & $3.8\times 10^{-7}$&8 & $2.2\times 10^{-7}$    \\  
1 & $5.0\times 10^{-7}$ &5 & $3.4\times 10^{-7}$ & 9 & $1.8\times 10^{-7}$   \\   
2 & $4.6\times 10^{-7}$ &6 & $3.0\times 10^{-7}$   &10 & $1.4\times 10^{-7}$  \\  
3 & $4.2\times 10^{-7}$  &7 & $2.6\times 10^{-7}$&&  \\ 
\end{tabular}
 \caption{Total decay widths of the intermediate states in a Li-He dimer, 
depending on the vibrational level. 
For fixed nuclei the decay width amounts to 
$\Gamma = 6.91\times 10^{-7}$\,a.u., ($\Gamma = 7.59\times 10^{-8}$\,a.u.) at $R=10$\,a.u. ($R=20$\,a.u.).}\label{tab: 1}
\end{center}\end{table}

Another relevant quantity is the integral resonance strength. It can be roughly estimated by the product of peak height and width. 
The corresponding value for $\nu_{\text{a}} = 0$ approximately is $\sigma^{(2)}\Gamma\big|_{\nu_{a}=0}\approx 1.34\times10^{-7}\text{ a.u.}$., for example,
whereas the {\it relative} integral resonance strength amounts to
$\eta^{(2)}\Gamma\vert_{\nu_{\text{a}}=0}\approx 0.65 $. Summing the relative integral resonance strength over all peaks, we obtain a total value of about $15 \text{a.u.}.$
It is comparable to values of $\approx 12 \text{a.u.}$ which are obtained in the atomic calculations $\eta_{\text{atomic}}(\omega_{B},R)*\Gamma(R)$ for a nuclear distance around $R_{\text{eq}}$. Hence, loosely speaking, the atomic resonance height is redistributed over the vibrational multiplet of resonance lines in the molecular case. This interpretation fits to our previous approximation in Eq. \eqref{eq: FC_Vereinfachung}.

In an experimental setup, the frequency of the external field is not exactly defined, but exhibits a width $\Delta\omega$. In order to account for this, we perform a corresponding averaging over all frequencies  by employing a Gaussian distribution with $\varsigma=\frac{\Delta \omega}{2\sqrt{2\ln{2}}}$. The averaged cross section reads
\begin{equation}
\bar{\sigma}^{(M)}(\omega)=\int d\omega' \sigma^{(M)}(\omega')\frac{1}{\sqrt{2\pi\varsigma^{2}}}e^{-\frac{(\omega'-\omega)^{2}}{2\varsigma^{2}}}.
\end{equation}
Applying an  FWHM energy width $\Delta \omega=1.7\text{meV}$ of the incident photon beam as used in  \cite{2CPIexp} we obtain averaged ratios depicted in Fig. \ref{fig:Mittelung}.
Due to the finite width of the field frequency, the peaks are reduced in height but also broadened. As a consequence, the integral resonance strength remains comparable to the one deduced from Fig. \ref{fig:P2/P1}. For example. when calculating the value again for $\nu_{\text{a}}=0$, we obtain the same order of magnitude as given above.

Before proceeding to the conclusions, we note that 
the inclusion of molecular rotations, which have been disregarded
in our treatment, is expected to lead to an additional fine structure
of the resonance lines in Fig. \ref{fig:P2/P1}. Each of them would be split into a multiplet of lines which are associated with different rovibrational
transitions from the ground to the intermediate state \cite{Fussnote_rot}. The resulting
line splitting within such a multiplet would be very narrow, 
since the energy differences between rotational states belonging to 
the same vibrational level is very small. An experimental resolution 
of this substructure would therefore be challenging. 

 \section{Conclusion and Outlook}
Resonant two-center photoionization in weakly bound systems has been studied, including the nuclear motion due to molecular vibrations. Earlier treatments of the process assumed spatially fixed nuclei and obtained, accordingly, a 'local' 2CPI cross section which strongly depends on the internuclear distance $R$ and exhibits a single resonance peak.
When treating the system in a molecular way instead, the $R$-dependent interaction potential is incorporated in a 'global' way by defining the vibrational states of the nuclear motion.

An approximate expression highlighting the molecular effects
was established (see  Eq. \eqref{eq: FC_Vereinfachung}) which shows that the changes
to the calculation assuming fixed $R$ can be attributed to
Franck-Condon overlaps and vibrational energy levels. As
a consequence, the two-center cross section splits up into
a multitude of peaks at various resonance energies which are
determined by the electronic and vibrational transition energies.
In the studied case of Li-He dimers, the cross section of each peak
is reduced as compared to the value obtained for fixed (equilibrium)
internuclear distance. Summing over all peaks, however,
yields a total value similar to the `local' resonant cross sections
which are obtained when a fixed internuclear separation $R$
close to the equilibrium distance is taken.

The latter observation indicates that theoretical treatments of
other two-center processes assuming fixed internuclear distances
can be expected to provide meaningful predictions for weakly bound molecular
systems, as well. Corresponding recent studies have considered,
for example, electron-impact ionization \cite{gruell} and photo double ionization \cite{Alex},
resonant electron scattering and recombination \cite{Eckey_Jacob}, and electron capture \cite{Sisourat}.
The influence of nuclear motion in a van-der-Waals molecule on (some of)
these two-center processes will be the subject of a forthcoming study.

\section*{Acknowledgement}
This work has been funded by the Deutsche Forschungsgemeinschaft (DFG, German Research Foundation) under Grant No. 349581371 (MU 3149/4-1 and VO 1278/4-1)



\begin{thebibliography}{33}
\bibitem{Fussnote2} Note that the relaxation of a two-center autoionizing state via ICD is a very fast process, whose timescale
was determined by various methods to lie typically in the pico- to femtosecond range; see  \cite{2CPIexp,Moiseyev,Jabbari,Fruehling}.
\bibitem{Fussnote} We point out that the width $\Gamma$ of the excited state in atom B
must be added by 'hand' here since we treat 2CPI in the lowest possible
order of perturbation theory. The origin of a non-zero width can be deduced in refined treatment
of the problem,
for example, by applying a nonrelativistic (noncovariant) QED approach
\cite{2CPI2} in which
both radiative and two-center Auger widths are shown to arise
naturally from the same
basic interaction between the electrons and the electromagnetic field,
or by summing parts of the perturbation series
to infinite order \cite{Kaspar}.
\bibitem {Fussnote_rot} In case of Li-He dimers, the ground state does not support rotational excitations \cite{Friedrich}.
\end{thebibliography}
\end{document}